\documentclass[reprint,aps,prl,nofootinbib,preprintnumbers,twocolumns,showpacs,10pt]{revtex4-1}
\usepackage{epsfig}
\usepackage{graphicx}
\usepackage{amsmath}
\usepackage{amsfonts}
\usepackage{amssymb}
\usepackage{epstopdf}
\usepackage{bm}

\newcommand\ee{\end{equation}}
\newcommand\be{\begin{equation}}
\newcommand\eea{\end{eqnarray}}
\newcommand\bea{\begin{eqnarray}}

\newcommand\vev[1]{{\langle {#1} \rangle}}

\newcommand\gev{\,\mbox{GeV}}
\newcommand\mev{\,\mbox{MeV}}

\renewcommand\({\left(}
\renewcommand\){\right)}
\renewcommand\[{\left[}
\renewcommand\]{\right]}

\begin{document}
\def\thefootnote{\fnsymbol{footnote}}
\title{Theoretical bounds on the tensor-to-scalar ratio in the cosmic microwave background}
\author{Lotfi Boubekeur}
\affiliation{
Abdus Salam International Centre for Theoretical Physics
Strada Costiera 11, 34151, Trieste, Italy.}

\begin{abstract}
Tensor modes in the cosmic microwave background are one of the most robust signatures of inflation. We derive theoretical bounds on the tensor fraction, as a generalization of the well-known Lyth bound. Under reasonable assumptions, the new bounds are at least two orders of magnitude stronger than the original one. We comment on a previously derived generalization, the so-called  Efstathiou-Mack relationship. We also derive a new absolute upper bound on tensors using de Sitter entropy bounds. 
\end{abstract}

\maketitle

\twocolumngrid
\section{I. Introduction\\[-3mm]}
Understanding the initial conditions that led to structure formation in our Universe is one of the most important issues in modern cosmology. Among the variety of available scenarios, inflation occupies a special place. In addition to being a theoretically attractive paradigm, it is compatible with all the available observational data.  In its simplest version, single field slow-roll inflation is realized through a  canonically normalized scalar field minimally coupled to Einstein gravity. To match with observations, the height and the slope of the potential have to obey special relationships. However, in most of the models, the excursion of the scalar field (inflaton) is at least of the order of the reduced Planck scale\footnote{Throughout the paper, we are using natural units $\hbar=c=1$.} $M_P=(8 \pi G_N)^{-1/2}$. Such large excursions potentially undermine the validity of effective field theory  used to derive the predictions. This is especially important considering the fact that quantum gravity corrections are likely to spoil the delicate balance between the height and the slope of the potential once the inflaton is allowed to travel over Planckian distances in field space\footnote{Notice that, in addition to spoiling the flatness of the potential, non-renormalizable operators $\lambda_n \phi^{n+4}/M_P^n$ (with $n\ge 1$) will make the energy density $\gg M_P^4$ once the inflaton takes super-Planckian values.}. Inflation, in this sense, is unique because it is doubly UV sensitive: first because of the mass of the inflaton owing to its scalar nature, and second because of the  unavoidable proliferation of dangerous UV-suppressed operators. Of course, supersymmetry addresses the first issue, however the fine-tuning of the infinite tower of Planck-suppressed operators inevitably remains. 

On the other hand, inflation predicts a stochastic background of gravitational waves (GWs) whose magnitude is related to the energy scale  of inflation \cite{Lyth:1984yz} and, more importantly, to the inflaton  excursion  \cite{Lyth:1996im}. According to the {\em Lyth bound} \cite{Lyth:1996im}, positive detection of tensors  would mean super-Planckian values of the inflaton which is clearly interesting both from the model building standpoint and from the observational one.  This  would also imply serious conceptual rethinking about effective field theory. Given the fact  that inflation offers an unequal and unique opportunity to access the Planck scale, and in the absence of experimental guidance on that question, it is important to use theoretical consistency to make some progress. In this article, we scrutinize this issue and derive theoretical bounds on the tensor fraction. 
\vspace{-4mm}
\section{II. The bounds\\[-3mm] } We begin by recalling  the necessary conditions under which the bounds  hold. First, we assume that gravity is described by Einstein general relativity (GR) and that inflation is of the slow-roll variety defined by the usual flatness conditions
\be
\epsilon\equiv{1\over2} M_P^2|V'/V|^2\ll 1\textrm{ and } |\eta|\equiv M_P^2|V''/V|\ll 1\; , 
\ee
together with the slow-roll approximations $3 H \dot\phi \simeq -V'(\phi)$ and $3 H^2 M_P^2\simeq V(\phi)$, where $H$ is the Hubble rate. Second, we assume that the primordial curvature perturbation $\zeta$, which is constant on superhorizon scales, is produced through the vacuum fluctuation of one or more light scalar fields. Using this approximation we can compute the spectrum of curvature perturbation produced by the inflaton ${\cal P}_\zeta(k)^{1/2}=H^2/2 \pi \dot\phi$. In the case of a single field, which we assume from now on, the latter  quantity should be equated with the observed value $\sim 5\times 10^{-5}$. The scalar index $n_s$ is given in the slow-roll approximation by $n_s= 1+2\eta_\ast -6 \epsilon_\ast$, where the subscript $\ast$ means  quantities are evaluated at horizon exit.  The latest WMAP 7 yrs dataset \cite{Komatsu:2010fb}
\be
n_s= 0.963 \pm 0.012\ \quad 68\% \textrm{ CL} 
\label{wmap}
\ee
excludes the Harrison-Zeldovich-Peebles spectrum by more than 3$\sigma$ and strongly favors a red tilt. On the other hand the tensor-to-scalar ratio $r\equiv 16 \epsilon_\ast$ is subject to the bound 
\cite{Komatsu:2010fb} $ r<0.24$ at  $95\%$ CL (WMAP+BAO+$H_0$). As we will see, these bounds are already quite constraining.  For instance, hybrid models based on $V(\phi)=V_0+m^2\phi^2/2$, with $m^2>0$, are  ruled-out because of their blue-tilted spectrum. Chaotic models based on $V\propto \phi^p$, with $p\ge 4$ are also ruled-out by this bound at least at the 3$\sigma$ level. We will focus hereafter on models that are still not excluded by WMAP, namely chaotic models with $p<4$ and hilltop models $V(\phi)=V_0-\lambda_n \phi^n$ with $n\ge 2 $.
\vspace{-4mm}
\subsection{A. Extended Lyth bounds\\[-3mm]} Our starting point to derive the bounds is the definition of the classical\footnote{By classical number of e-folds, we mean $N$ in the slow-roll noneternal inflation regime. } number of e-folds
\be
d N= M_P^{-1}  {d \phi /\sqrt{2 \epsilon(\phi)} } \, . 
\label{N}
\ee
The total number of e-folds is obtained as usual by integrating Eq.~(\ref{N}) starting from horizon exit to the end of inflation. The original bound \cite{Lyth:1996im} was derived by integrating Eq.~(\ref{N}) for modes corresponding to the multipoles $2< \ell\lesssim 100$. The crucial point is that $\epsilon$, and thus $r$, does not change too much during the last $\Delta N \simeq 4$ e-folds corresponding to these modes. Under this reasonable assumption,  Eq.~(\ref{N}) gives   
\be
{\Delta \phi_{4}\over M_P}\simeq \(r\over 0.52\)^{1/2}\, ,
\label{lyth}
\ee
where $\Delta\phi_{4}$ is the inflaton displacement during the last 4 e-folds.  However, if one wants to extend that bound to the whole $N$ e-folds, one {\em cannot} assume  negligible variation of the slow-roll parameters anymore. In the following, we will derive bounds on the inflaton    excursion taking into account the variation of $\epsilon$ during inflation. Before doing so, we will derive a general inequality which is valid for all slow-roll models. Using the fact that the number of e-folds $N$ is just the area under the curve $(1/M_P \sqrt{2 \epsilon(\phi)})$ between $\phi_\ast$ and $\phi_{\rm end}$, we can write that\footnote{\label{fn}We are using the following basic property of definite integrals: if a function $f(x)$ is bounded on an interval 
$a\le x\le b$ i.e. $A \le f(x)\le B$ then $(b-a) A\le \int_a^b dx~f(x)\le  (b-a) B$.}
\be
{\Delta\phi\over M_P} {1\over \sqrt{2 \epsilon_{\rm max}}}\le N\le
{\Delta\phi\over M_P}{1\over \sqrt{2 \epsilon_{\rm min}}}\;  ,
\label{boundN}
\ee
where $\Delta\phi\equiv |\phi_{\rm end} - \phi_\ast |$ is the total inflaton excursion during $N$ e-folds and $\epsilon_{\rm min}$ and $\epsilon_{\rm max}$ are the minimum and maximum values of $\epsilon$ during that period. Eq.~(\ref{boundN}) is the first main result of this paper. In principle, $\epsilon_{\rm min}$ could be vanishingly small; however, for the validity of the semiclassical description, the smallest possible value, which corresponds to the phase transition to the eternal inflation regime \cite{Creminelli:2008es}, is $\epsilon_c \equiv 3 (H/M_P)^2/4 \pi^2$. Plugging in this value, we obtain $N\lesssim \Delta\phi/H$. Typically, though, the slow-roll parameter grows monotonically during inflation from $\epsilon_\ast$ until the breakdown of slow-roll $\epsilon_{\rm end}\sim 1$. This is, in fact, the case for hilltop and chaotic inflation scenarios to which we will apply the bounds,  Eq.~(\ref{boundN}). 

Let us begin with hilltop inflation models where $\epsilon$ increases monotonically as the Universe inflates. Using the right-hand side of Eq.~(\ref{boundN}), we get a bound that can be written in terms of the tensor-to-scalar ratio as
\be
r<0.002 \({\Delta \phi \over M_P}\)^2 \(60 \over N\)^2\,.
\label{boundr}
\ee
Notice that Eq.~(\ref{boundr}) is stronger than the original bound derived from Eq.~(\ref{lyth}) by a factor of $\sim 230$. It is also stronger than the variant derived in \cite{baumann} by a factor 4. This discrepancy is due to the fact that data at the time allowed for a significantly larger negative value for $\tilde\eta\equiv d \ln \epsilon /d N$. As a result, relatively low values $N_{\rm eff}\simeq 30$ of the effective number of e-folds defined  in Eq.~(5) of \cite{baumann} were  allowed. This is no longer the case in light of present observations. In particular, according to recent SPT results \cite{spt}, $\tilde\eta>0$ and $N_{\rm eff}\gtrsim 88$. 

Now, let us focus on quadratic hilltop models whose potential is given by $V(\phi)=V_0- m^2 \phi^2/2 + \cdots$ with $m^2>0$. The dots stand for higher order terms that will make the potential bounded from below and which might dominate after horizon exit. In this case, the tensor-to-scalar ratio is given by\footnote{ 
If the potential does not steepen after horizon exit, the tensor-to-scalar ratio  will satisfy a more relaxed bound \cite{Lyth:1996im}
\be
r=8(1-n_s) \, e^{-N(1-n_s)}\lesssim 0.03\, (60/N)\,.
\ee
}
\be
r=2(1-n_s)^2\, \({\phi_{\rm end}\over M_P}\)^2 e^{-N(1-n_s)}\, . 
\label{bound}
\ee
From Eq.~(\ref{bound}),  we can use the WMAP 7 yrs bound on the scalar spectral tilt $1-n_s\lesssim 0.04$ to derive a more stringent bound on $r$ which reads\footnote{One can use the property that $a^n \, e^{-a}<b^n e^{-b}$ for any $a>b>n\ge 0$.  } \cite{Boubekeur:2005zm} 
\be
r\lesssim 0.0003 \({\Delta \phi\over M_P}\)^2 \(60 \over N\)^2\, , 
\ee
which is one order of magnitude stronger than the previous bound.

One can also consider variants of the hilltop scenario where higher powers of $\phi$ dominate at the top of the potential. If there is no $\phi\to -\phi$ symmetry then the first term in the potential will be cubic. However, and in addition to the fact that the potential is not bounded from below, this leads to a spectral index on the verge of the 3$\sigma$ WMAP allowed region. If on the other hand there is a symmetry $\phi\to-\phi$ and the quadratic term is, for some reason,  negligible\footnote{Notice that, apart from shift symmetries $\phi\to\phi+c$,  there is no symmetry that can consistently forbid the quadratic term.}, then the potential will be $V(\phi)=V_0-\lambda \phi^4/4$. If inflation is responsible for the generation of density perturbations then $\lambda\simeq 10^{-12}$ regardless of the value of $V_0$. This tiny value of the coupling can be justified if one thinks of $\phi$ as a modulus described by the potential $V_0(1-\lambda_4\phi^4/M_P^4)$ with $\lambda_4\sim 1$ and $V_0\sim 10^{15}\gev$. In the following we will consider general values of $V_0$, allowing both sufficient reheating and the use of standard QFT and GR methods $\mev^4 \ll V_0\ll M_P^4$. The bound Eq.~(\ref{boundr}) still applies in this case; however, one can derive a tighter relationship between $\Delta\phi$ and $r$ which can be written as
\be
{\Delta\phi\over m_P}={N^{3/4}\over 2 \sqrt{\pi} } \, r^{1/4}\,, 
\label{EM}
\ee
where $m_P=G_N^{-1/2}$ is the Planck mass. For $N=60$, Eq.~(\ref{EM}) gives, indeed, the Efstathiou-Mack relationship  $\Delta\phi/m_p\approx 6 r^{1/4}$ which was derived empirically in \cite{Efstathiou:2005tq}. Actually, Eq.~(\ref{EM}) is a special case of a more general relationship, \footnote{Here, we are assuming reasonably that $\phi_*\ll \mu$.  } 
\be
{\Delta\phi\over m_P}=r^{p-2\over2p} \frac{\[p(p-2)N\]^{1/p} }{\sqrt{8 \pi}} \({N(p-2)\over 2 \sqrt{2}}\)^{p-2\over p}\, , 
\label{family}
\ee
which holds for general hilltop models described by the potential $V(\phi)=V_0\[ 1-\lambda_p (\phi/\mu)^p\]$, where $p>2$ and $M_P>\mu>0$ regardless of the values of the parameters of the potential i.e. $V_0$,  $\mu$ and $\lambda_p$. Therefore, there is a whole family of models that  satisfy this relationship. 
Next, let us consider chaotic models which are characterized by a power-law potential $V(\phi)\propto \phi^p$. As hilltop models they have the property that $\epsilon$ increases monotonically during inflation.  Therefore, the bound on the tensor-to-scalar ratio Eq.~(\ref{boundr}) still holds as in the case of hilltop inflation. However, replacing $\phi_\ast\simeq \sqrt{2pN}M_P$, the resulting bound  
\be
r\lesssim 0.27 \, p \(60 / N\)\; .
\ee
is hardly constraining even for $p=1$. 

What about Natural inflation? Natural inflation is the only know field theoretically consistent implementation of chaotic inflation. The  inflaton is a PNGB, with a symmetry breaking scale $f\gg M_P$. In general, the potential can be written as $V(\phi)\propto [1+ \cos(\phi/f)]$, where $f\equiv M_P/\sqrt{2|\eta_0|}$; $\eta_0<0$ is the second slow-roll parameter at the top of the potential, and it reduces to a quadratic hilltop potential for small angles. For $N=60$, using the WMAP 7 yrs bound on $n_s$, we can constrain (See e.g. Sec. 6 of \cite{Boubekeur:2005zm} for details) $|\eta_0|\lesssim 0.017$. This, in turn, gives a range of tensor fraction $0.04 \lesssim r\lesssim 0.13$, that is compatible with the later bound. 
\vspace{-4mm}
\subsection{B. de Sitter entropy bounds\vspace{-3mm}}Let us now consider de Sitter entropy bounds. The second law of thermodynamics states that the entropy of any closed system never decreases with time. The expanding Universe during slow-roll inflation does not escape this rule \cite{Gibbons:1977mu}. The entropy of de Sitter spacetime is given by $S_{dS}=8\pi^2 M_P^2/H^2$, and it varies during $N$ e-folds of slow-roll as 
\be 
{dS_{dS}\over dN}=16 \pi^2 {M_P^2 \,\epsilon \over  H^2}=2 \vev{\zeta^2}^{-1}\,, 
\label{s}
\ee
where we used $\vev{\zeta^2}\equiv H^2/ 8 \pi^2 M_P^2 \epsilon$. The second law in this case boils down to the requirement $\dot H<0$ during inflation. It is easy to show that the integral of Eq.~(\ref{s}) is bounded as follows 
\be
{ \langle\zeta_{\rm max}^{2}\rangle}^{-1}\le \Delta S_{dS}/2 N\le
{ \langle\zeta_{\rm min}^{2}\rangle}^{-1}\, .
\label{entropy}
\ee

\begin{figure*}[!t]
\vspace{-0.6cm}
\begin{center}
\includegraphics[width=0.73\textwidth]{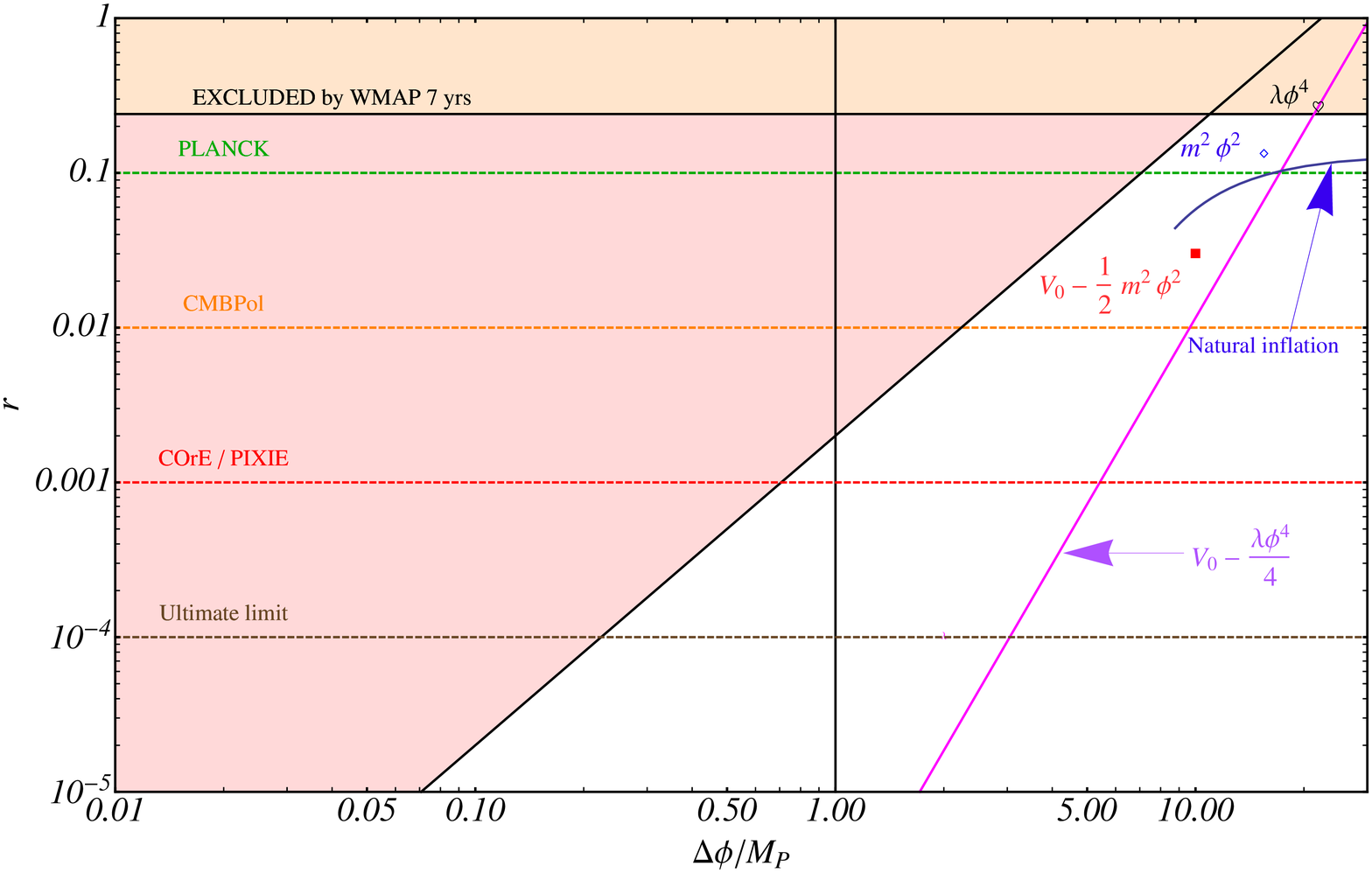} 
\caption{The tensor-to-scalar ratio versus field excursion for inflationary potentials compatible with WMAP 7 yrs result on the spectral tilt Eq.~(\ref{wmap}). The light orange region is excluded by WMAP 7 yrs bound on tensors.  The black diagonal line represents our bound Eq.~(\ref{boundr}), the pink area is excluded. The magenta diagonal line represents the Efstathiou-Mack relationship \cite{Efstathiou:2005tq} describing models based on the potential $V(\phi)=V_0-\lambda\phi^4/4$. The black vertical line corresponds to $\Delta\phi=M_P$. The horizontal  dashed lines represent the forecasted sensitivity of forthcoming observations. All the models are plotted for $N=60$.}
\label{fig:bounds}
\end{center}

\end{figure*}

This inequality is valid independently of the variation of the slow-roll parameter in the noneternal inflation regime. Neglecting the Hubble parameter variation during inflation and taking 
$\epsilon_{\rm min}=\epsilon_c$ for noneternal inflation the left-hand side of Eq.~(\ref{entropy}) yields the well-known bound on e-folds  from de Sitter entropy \cite{Dubovsky:2008rf}, $N\le S_{dS}/12$. On the other hand, the right-hand side of Eq.~(\ref{entropy}) leads to the trivial and model-independent lower bound on the number of e-folds,  
$N\ge 1/2$. Now, specializing to the typical case of monotonically increasing $\epsilon$, i.e. $\epsilon_{\rm min}=\epsilon_\ast$,  the left-hand side of Eq.~(\ref{entropy}) yields $N\le { \langle\zeta_\ast^{2}\rangle} \Delta S_{dS}/2$, which, in turn, can be written as a general model-independent bound on the tensor fraction  
\be 
r\lesssim 0.13\,  (60/N)\, \[(H_\ast/H_{\rm end})^2-1\]. 
\label{boundent} 
\ee
This is the second main result of this paper. Due to the prefactor $(H_\ast/H_{\rm end})^2$, Eq.~(\ref{boundent}) is more stringent for models where $H\simeq$ constant. On the other hand, once $r$ is measured, one can straightforwardly turn Eq.~(\ref{boundent}) into an upper bound on the Hubble rate at the end of inflation.  
%
\vspace{-4mm}
\section{III. Observational constraints\\[-3mm]}Let us now discuss the observational situation which is really promising. The {\sl Planck} mission \cite{PLANCK} is aiming to reach $r \lesssim 0.1$ in which case the simplest quadratic chaotic inflation scenario will be probed. In addition, a variety of experimental setups \cite{CMBPOL, CORE, Kogut:2011xw} are targeting the range $10^{-3} \lesssim r \lesssim 0.1$.  The lowest detectable tensor fraction through CMB polarization is probably $r\simeq 10^{-4}$ both from polarized dust foregrounds substraction \cite{lowestforgrounds} and contamination from $E$ to $B$-modes conversion through lensing \cite{lowestlensing}. In Fig.~\ref{fig:bounds}, we represent the various models in the plane $\Delta\phi$ versus $r$, instead of the traditional $r$ versus $n_s$ plot, together with our bound Eq.~(\ref{boundr}), and experimental reaches of the planned observations. We do not represent the family of models that satisfy Eq.~(\ref{family}) as they lie close to the magenta line representing the Efstathiou-Mack relationship. As expected, Natural inflation appears to interpolate between quadratic hilltop and quadratic chaotic models. It is also noteworthy, though well known, that none of the single-field scenarios compatible with observation has a sub-Planckian inflaton excursion.
\vspace{-5mm}
\section{IV. Conclusions\\[-3mm]}In conclusion, we reiterate the truism that the tensor fraction is a very important observable to test the initial conditions of the Universe. We derived two theoretical bounds on that quantity. The first bound Eq.~(\ref{boundr}) means that (for sub-Planckian inflaton excursion and thus consistent field theory description) $r\lesssim 0.002$, placing it beyond the reach of {\sl Planck} but within reach of {\sl COrE} and {\sl PIXIE} \cite{Kogut:2011xw}. On the other hand, the fact that both single-field benchmark scenarios that are consistent with current data have  $\Delta\phi \gtrsim 10\, M_P$ make their tensor fraction  within reach of {\sl CMBPol}. The second bound,  arising from de Sitter entropy bounds,  implies an upper bound on tensors Eq.~(\ref{boundent}), independently of the magnitude of inflaton excursion, which is within reach of {\sl Planck} and future observations. The detection of $B$-modes would promote inflation from an attractive paradigm to a predictive theory. However this is not the end of the story, as any realization of slow-roll inflation will have to face the issue of a consistent UV-completion.  

We end up with some speculations. It would be interesting to explore the relationship between the entropy of the inflaton and $r$. For instance, in \cite{Conlon:2012tz} it was argued that the entropy of the inflaton is proportional to $(\Delta\phi/H)^2$, which by the de Sitter entropy bound,  would censor any attempt to have super-Planckian inflaton excursion, and thus observable inflationary GWs, in a consistent UV-complete theory. This is in agreement with the conclusions reached in \cite{Banks:2003sx}, where it was shown that attempts to build natural inflation models in string theory with a decay constant $f\gg M_P$ are doomed to failure. 
\vspace*{-5mm}
\section{Acknowledgments\\[-3mm]}
I thank Paolo Creminelli and Scott Dodelson for providing invaluable comments on a previous version of the draft. I am also grateful to David Lyth for initial collaboration and inspiring this project.

\onecolumngrid

\end{document}